\documentclass[twocolumn]{article}

\usepackage{arxiv}
\usepackage{graphicx}
\usepackage[utf8]{inputenc} % allow utf-8 input
\usepackage[T1]{fontenc}    % use 8-bit T1 fonts
\usepackage{hyperref}       % hyperlinks
\usepackage{url}            % simple URL typesetting
\usepackage{booktabs}       % professional-quality tables
\usepackage{amsfonts}       % blackboard math symbols
\usepackage{nicefrac}       % compact symbols for 1/2, etc.
\usepackage{microtype}      % microtypography
\usepackage{lipsum}

\title{Machine learning and chord based feature engineering for genre prediction in popular Brazilian music}

\author{
  Bruna D. Wundervald\\
  Department of Statistics/Hamilton Institute\\
  Maynooth University\\
  \texttt{brunaaviesw@gmail.com} \\
   \And
 Walmes M. Zeviani \\
  Department of Statistics\\
  Paraná Federal University\\
  \texttt{walmes@ufpr.br} \\
}

\begin{document}

\twocolumn[{%
\maketitle
  \begin{@twocolumnfalse}
    \maketitle

\begin{abstract}
Music genre can be hard to describe: many factors are involved, such as style, music technique, and historical context. Some genres even have overlapping characteristics. Looking for a better understanding of how music genres are related to musical harmonic structures, we gathered data about the music chords for thousands of popular Brazilian songs. Here, 'popular' does not only refer to the genre named MPB  (Brazilian Popular Music) but to nine different genres that were considered particular to the Brazilian case. The main goals of the present work are to extract and engineer harmonically related features from chords data and to use it to classify popular Brazilian music genres towards establishing a connection between harmonic relationships and Brazilian
genres. We also emphasize the generalisation of the method for obtaining the data, allowing for the replication and direct extension
of this work. Our final model is a combination of multiple classification trees, also known as the random forest model. We found that features 
extracted from harmonic elements can satisfactorily predict music genre for the Brazilian case, as well as features obtained from the Spotify API. The variables considered in this work also give an intuition about how they relate to the genres.

\end{abstract}

  \end{@twocolumnfalse}
}]

% keywords can be removed

\section{Introduction}

Many factors are involved in the configuration of a music genre: style, music techniques, historical context, regional identity, harmonic structures
and so on \cite{Caldas2010}.
A list of popular music genres can be easily found, but it usually does not come with no proper description for each one \cite{Lena2008}. For the Brazilian case, 
some music genres even hold overlapping characteristics, 
making it harder to distinguish one from another.
Musical genres facilitate the search for music, 
enveloping a simplification of what each song
is. \cite{Lee2004}
found that users prefer
to look for music of their interest by 
genre rather than by other metrics, such as
artist similarity. Thereby, inconsistencies
and blurriness in the definition of musical
genres pose an important problem in 
various aspects of music studies. 

Meanwhile, as \cite{Cheng2008} 
and \cite{Correa2016}
observed, mid-level
music features such as chords configure a rich
resource of information regarding genres. The
chords sequence of a song fully describes its
harmonic progression and, by consequence, represents
a meaningful part of the total music structure.  
Therefore, the focus of this work is to establish
a connection between harmonic information 
and genre specification in Brazilian popular music. 
Here, the term 'popular' does not refer only to 
the genre named MPB, that stands for 
\textit{Música Popular Brasileira}
\cite{albin2003livro}, or
Brazilian Popular Music, but to nine 
genres that were considered familiar to the
Brazilian context.

Musical data is available in a wide variety of formats
for the genre classification task: music sheets, 
chords, lyrics, MIDI, audio files, and 
others \cite{Burgoyne2015}. Each one of those 
formats carries different levels of information 
about the pieces. The choice of a data format to 
work also defines what kind of information will 
be available for the analysis. The 
choice of symbolic chords data allows
us to extract harmonically related features
for the genre classification. More than
just using raw information, we can capture deeper
characteristics of the songs by 
representing their chords structures in
different and meaningful forms. 

Related work has been done for most of the usual
representations of music data. For audio data,
we can mention \cite{Pampalk}, 
\cite{Tzanetakis}, 
\cite{Bahuleyan} and 
\cite{Scaringella}, all of which focused
in music genre classification using
audio extracted features. Concerning
text data related to music,
\cite{Neuman2016} presents a discussion
about the characterization of genres
through song lyrics. A recent interesting publication is \cite{Chuan2018}, where 
the authors introduced a vector 
based representation for chords sequences.
This work is notable in the sense that it
brings to light a new and effective way
to extract information about symbolic 
chords data. A similar problem to ours was attacked in \cite{PerezSancho2010},
where the authors also focused on harmonic features for genre classification. 
However, only three musical genres were considered,
while the role of chords in the classification 
was given only by its symbolic form, 
making it overly simplistic and leaving room
for a better representation of the
chords structures. Furthermore,  in all of the 
mentioned, researchers did not focus
neither on the manual extraction
and consequent interpretation 
of the features or on extending the method
used for the obtention of the data, which
are the premises of our approach.

It is also important to notice that, until recent times,
studies about popular music in Brazil were
conducted typically by folklorists and musicians
who 
worked mainly with classic Brazilian 
genres, such as Samba and Bossa Nova 
(\cite{Vianna1999}, \cite{Behague1980} and
\cite{McCann2007}). Music research was more 
focused on tradition and ethnic matters of Brazilian culture and their authenticity against foreign music. 
During the last few years, 
research in Brazilian popular music 
has become stronger and more inclusive concerning 
genres, but
we still lack references when looking for 
works within the machine learning context. 
Recognizing that music information retrieval 
is an impactive study field \cite{Muller2007},
we see this as an opportunity to bring the
two subjects together. 

The main goals of the present work are
to extract and engineer harmonically related
features from chords data and use it
to correctly classify popularly
Brazilian music genres, towards
establishing a connection between harmonic
relationships and music genres.
We also emphasize on the generalization 
of the method for obtaining the data, allowing
for the replication and direct extension
of this work. The
following section describes how the data 
was obtained and analyzed, being complemented
in Section 3 by the methods for the music genre
modeling. We present the results 
regarding the construction of the variables,
exploratory analysis and machine learning
models applied in Section 4. Conclusions 
are in Section 5, containing a
summary of the results, future work and 
final remarks about the project. 

\section{MATERIALS}\label{sec:page_size}

A musical chord is a grouping of three or more notes
played simultaneously \cite{Almada2012}. There is usually 
a third interval between each note. When the chord
is composed of three notes we have a triad, 
and when it has four notes, a tetrad. 
Triads are formed by a fundamental note, a third and a fifth note, being that a tetrad is different
because it is added by a seventh note. 
These simple structures can be altered in many different ways, either by changing the current,
removing or adding notes. Other nomenclatures
rise depending on the change made, namely major, minor, augmented and diminished chords. 

The same chords can have different tonal functions, 
which are determined by their positions (or degrees) in the 
underlying music scale of a song. For example, the C major chord
is the first degree in the C scale, characterizing it
as the tonic chord, the one that gives the human ear a 
sensation of closure. On the other hand, the C major 
chord has a totally different function when played within
the F scale, where it assumes the dominant role and 
carries a tension feel. 

Chords progressions are a sequence of chords that happen
in a particular form. Some chords
progressions have familiar patterns and   
it is well known in harmony theory that 
the occurrence of a chord is correlated to the 
previous one for a reason \cite{Almada2012}. 
The most common case of this dependence is when a 
\textit{tension chord} happens and it is 
followed by a \textit{resolution chord}, 
giving the human ear a feeling of
satisfaction after the tense moment. 
Even with the existence of patterns, those structures 
can actually be very diverse, inducing a wide range
of sensations \cite{Rowe2005}. 
The way how a progression of 
chords is organized defines how the harmonics of a
song is structured and different genres explore it
in their own way. 

\subsection{Data Extraction}\label{subsec:data}

The data was extracted through \textit{webscraping} techniques 
\cite{Iacus2015},
from the Cifraclub website \\ 
(https://www.cifraclub.com.br/), an online
collaborative Brazilian page of music-sharing.
The website has a big collection of music chords 
for different instruments, and it is colaborative
in the sense that most of the chords information
present there are contributions of the users, 
what confers to the extracted data a natural
high variability. 

In total, nine music genres were used: 
Reggae, Pop, Forró, Bossa Nova, Sertanejo, MPB, 
Rock and Samba, chosen for being considered 
good representatives of the Brazilian music,
since they are constantly present in the dynamic
'Top Brasil' Spotify playlist \cite{Schettino2017}. 
Also, all of the nine genres appear on the main page of 
the Cifraclub website, where the most popular genres are. 
Other famous genres, as the Brazilian Funk and
electronic music were,
not evaluated as they can be better described by
their percussion characteristics instead of the 
harmonics. From the selected genres, 106 
different artists were available (solo singers or bands) on the Cifraclub platform. 

Our extraction method consists of the access of the source code of each 
song for all the artists and, from the 
underlying \textit{html} structures of the web pages,
the collection of chords, keys, name
of the artist and the name of the songs. 
Complementary variables about the release year and popularity were obtained with the aid of the Spotify \textit{API}.  We finished with a dataset of
almost 484.000 rows related to 106 artists and 
8339 different songs from the 9 Brazilian music 
genres.

To amplify the coverage of this work, the data extraction process resulted in the \textit{chorrrds} package ~\cite{chorrrds}, for the statistical software 
\texttt{R} \cite{Rsoftware}. 
The package is available for download and use from \textit{CRAN}\footnote{https://cran.r-project.org/web/packages/chorrrds/index.html}, the official repository of \texttt{R} packages. We point out that the creation of the package is especially beneficial for the
reproducibility purposes of this analysis, and 
for ensuring that the method can be generalized and
even extended. 

\subsection{Feature Engineering}\label{subsec:feat}

One important step of modeling is
feature engineering \cite{featureeng},
that can directly influence the
effectiveness of a statistical model. 
Feature engineering techniques are used
when transformations of variables into features
that represent the adjacent problem in a better way are needed \cite{nargesian2017learning}, which
depends on the context of the problem. The usefulness 
of the method for this work is explicit once the 
symbolic representation of chords progressions 
do not capture all of the possible useful information. 

If we account for the chords in their simplest form,
ignoring the possible alterations,
it is noticeable that the same chord progressions 
inevitably turn up in some music genres, for example.
Out of that, we consider that using only the
raw information about the chords progression
may cause a lot of information to be lost in the 
process. The details, such as if the chord was
changed in some way or how prevalent they are
in a song, should be taken into consideration. 
We emphasized on obtaining various distinct 
features from the chords, being able to make 
use of more information than only their symbolic form.

Feature engineering is typically automatic
\cite{nargesian2017learning}, meaning that
the model learns the features without much human intervention. These features
are hard to interpret or the interpretation 
is out of the scope of the research. 
Here, an opposite approach was employed,
We chose to employ a hand-engineering approach
to the variables, as we are interested
in obtaining features that are interpretable
in accordance to the posed problem. 
The resulting features will not only help 
with the modeling problem, but also
have a clear interpretation of 
harmonic characteristics of the considered
songs. In Table \ref{tab:ft}, a small demonstration 
of how the features were constructed is presented.

\begin{table}[ht]
\caption{Example of how features were manually extracted.} 
\[
  \begin{array}{|c|c|c|c|}

      chord & major
    & contain \thinspace 7th
    & contain \thinspace 6th \\
    \hline
    C   & 1 & 0 & 0 \\
    Gm7 & 0 & 1 & 0 \\
    \hline
    \multicolumn{1}{c}{} &  \multicolumn{1}{@{}l@{}}{%
      \raisebox{.5\normalbaselineskip}{%
      \rlap{$\quad \quad\underbrace{\hphantom{\mbox{$p\lor{}q$\hspace*{\dimexpr11\arraycolsep+\arrayrulewidth}$p\land{}q$}}}_{extracted\thinspace features}$}}%
    }
  \end{array}
\]
\label{tab:ft}
\end{table}

The technique allowed us to have a dataset
composed of 23 variables related to the songs,
detailed in the next section
of this paper. These features were divided into four
thematic groups: triads (6),
tetrads (6), common transitions (3) and miscellany (8). 
As the objective is not to model the songs but 
the music genres in which they are classified,  
the features went through a summarizing process.
For example, consider the second column of 
Table \ref{tab:ft}. This column shows an indicator
variable of the current chord being major
or not. For each song, this column was turned into the percentage of major chords. The
same was adequate for the last two columns of the
example table. In this fashion, we summarized 
our extracted variables, ending
up with one row of information per song. 

\section{MUSIC GENRE MODELING}\label{sec:model}

In this section, we discuss how the variables
were extracted and the method used for the 
music genre modeling.

\subsection{Extracted and Engineered Variables}

As mentioned before, one of our main
concerns is that the new features
can be interpretable. The engineered variables 
were separated into four thematic groups,
organized as follows:  

\begin{itemize}
    \item First group  - Triads and simple tetrads: 
\begin{enumerate}
\setlength{\itemsep}{0.2pt}
\item Percentage of suspended chords (e.g. Gsus).
\item Percentage of chords with the seventh  (e.g. C7).
\item Percentage of minor chords with the seventh  (interaction between the minor third and the seventh) (e.g. Em7, C\#m7).
\item Percentage of minor chords (e.g. Em, C\#m).
\item Percentage of diminished chords (e.g. Bº).
\item Percentage of augmented chords (e.g. Baug).
\end{enumerate}

\item Second group  - Dissonant Tetrads: 
\begin{enumerate}
\setlength{\itemsep}{0.2pt}
\item Percentage of chords with the fourth (e.g. D4).
\item Percentage of chords with the sixth (e.g. E6).
\item Percentage of chords with the ninth  (e.g. G9).
\item Percentage of minor chords with the major seventh 
(e.g. F7+, Am7+).
\item Percentage of chords with a diminished fifth (e.g. C5- or C5b).
\item Percentage of chords with as augmented fifth (e.g. C5+ ou C5\#).
\end{enumerate}

    \item Third group  - Main Chord Transitions:
\begin{enumerate}
\setlength{\itemsep}{0.2pt}
\item Percentage of the first most common chord transition in the song. 
\item Percentage of the second most common chord transition in the song. 
\item Percentage of the third most common chord transition in the song. 
\end{enumerate}

    \item Fourth group  - Miscellany:

\begin{enumerate}
\setlength{\itemsep}{0.2pt}
\item Popularity of the song, extracted from the Spotify
API.
\item Total of non-distinct chords in the song. 
\item Year of album release, extracted from the Spotify
API.
\item Indicator of the key of the song being the same
as the most common chord.

\item Percentage of chords with varying bass
(e.g. C/E, C/G, C/Bb).

\item Mean distance of the chords to the 'C' chord
in the circle of fifths. 

\item Mean distance of the chords to the 'C' chord
in semitones.  

\item Absolute quantity of the most common
chord.

\end{enumerate}
\end{itemize}

The first group contains information about small 
changes that can be done in the fundamental state of
a chord. It captures when the song is the result
of slight alterations in the chords. 
The second comprises more unusual
extensions, commonly made by experienced musicians. 
By forming these groups of variables
we can expect that if
differences in harmonic structures
between genres are really relevant, 
those characteristics would be pertinent to 
the harmonic characterization of the genres. 

The set of transition variables is also 
organized with the intention of capturing the
refinement of harmonic structures. For example, 
simplistic pieces tend to have a lesser variety of 
chords, but this does not mean that the songs last
less time. This implies that the same group of chords
and their transitions will be repeated many 
times for the whole duration of the song. 
In consequence, the top most common transitions 
would have high percentages, since there are not 
numerous different transitions happening in the song.
A new dimension of complexity of the songs is 
mapped in this group of features. 

The last group is a complete miscellany. 
It mixes very dissimilar predictors, 
possibly with distinct interpretations 
to the genres. We still have some features 
that concern the harmonic structures, 
in case they were not captured before,
including the total of chords, the percentage of chords with varying bass and mean distances
of the chord to the circle of fifths and by
semitones. To explain a little, 
the first variable contains
the information about the duration of the song,
while the varying bass is also indicative 
of a more complex piece. 

Besides that, in music theory
relationships between chords can be measured with the
circle of fifths \cite{Lewin1982}. 
To contextualize, this a circle that 
represents music chords, each one being  
a fifth apart from its neighbor. Usually, the 
neighbors appear together in a song,
as their distance in a circle of fifths
works as an indicator of their
harmonic functions. For example, 
when the C chord is been played, it is
very likely to be followed by the G, which
is its adjacent neighbor in the circle
of fifths \cite{Lerdahl1977}. We made use of 
this meaningful representation to obtain the 
distance features from our data. By calculating the
mean distances in the circle of fifths for each 
song, we can estimate how the variations of 
chords happen in the selected songs.
We already know from  \cite{Absolu2010}
that some pieces do not have highly varying
progressions, while others experiment with 
particular combinations. As for the last features, 
the year has the temporal dimension and popularity 
is composed of the public opinion about the songs,
which is as external element. These
two last features were extracted from the Spotify API. 

Starting only with the 
symbolic representation of the chords, we gathered
a broad set of information about the considered pieces.
A fundamental characteristic
of this data is the interpretation behind each feature.
Since one of our goals is to better understand how
genres are related to musical harmony in general, 
the interpretation is a primary element of this work. 
The variables were complemented with selected
features extracted from the Spotify API that carry important details. 

\subsection{Trees and Random Forests}

First, consider a variable of interest $Y_i \in \mathbb{R}$ 
and $\mathbf{x} = (x_{i1},\dots, x_{ip})'$ the set 
of predictor features, $1 \leq i \leq p$. A statistical
framework for non-parametric regression characterizes
their relationship as 

\begin{equation}
y_i = f_0(\mathbf{x_i}) + \epsilon_i, \thinspace
\epsilon_i \stackrel{iid}\sim N(0, \sigma^2),
\end{equation}

where $f_0$ is an unknown regression function. 
A regression tree is a completely non-parametric model,
described graphically by a tree,
that can be used to reconstruct $f_0$ by mapping the
predictors into a set of decisions at each step. 
On the literature, trees are applied in decision problems where there is no stochastic assumption made,
but instead, there is interest in finding 
rules that would lead to the right predictions \cite{Faraway2016}. 
Each rule has the form: $x_j > x_{j,th}$,
where $ x_j$ describes the value of the feature at $j$ and $ x_{j,th}$ is the decision cut point
\cite{Kingsford2008}. 
For the purpose of explaining how the trees work, consider a classification task for a factor target variable.  In our case, this factor variable is
the music genres. At the start, the full training dataset is
all in just one node of the tree. In each step of the algorithm, a binary partition is made in the covariables
space, meaning that a split results in two children that
separate better the target variable in its classes. The 
usual criteria to choose the best threshold for a new node is the Gini impurity \cite{Hastie}, measured by

\begin{equation}
Gini = 1 - \sum_{c = 1}^{r} p_c^{2},   
\end{equation}

where $p_c$ is the observed proportion for each class $c = 1,..r.$ in the dataset. This measure has its minimum when the individuals belong to the same class. Thereby, 
partitions are made in order to minimize the difference between the impurity, or heterogeneity, of the current state of the model and the next step with the candidate new node. This difference is written as

\begin{equation}
\Delta Imp = Imp_{O} - \Big( \frac{n_1}{n_{O}}
Imp_{1} + \frac{n_2}{n_{O}} Imp_{2},
\Big)
\end{equation}

where $1$ e $2$ are the candidate divisions, starting at region $O$. Consequently, 
the algorithm looks for the splitting
cut point that leads to the minimal
combined impurity of the groups. 

\begin{figure}[t]
\centering
\includegraphics[width=0.6\columnwidth]{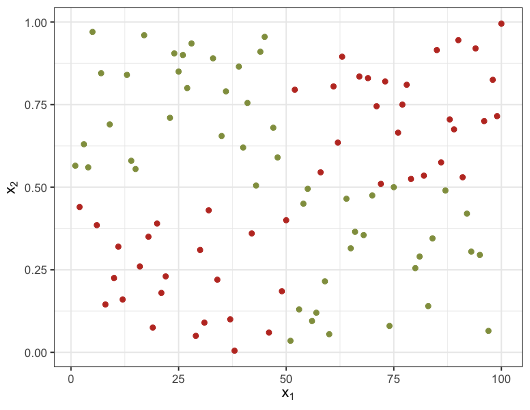}
\caption{Two covariables with the target variable groups mapped in the color of the points.}
\label{fig:tree}
\end{figure}

\begin{figure}[t]
\centering
\includegraphics[width=0.6\columnwidth]{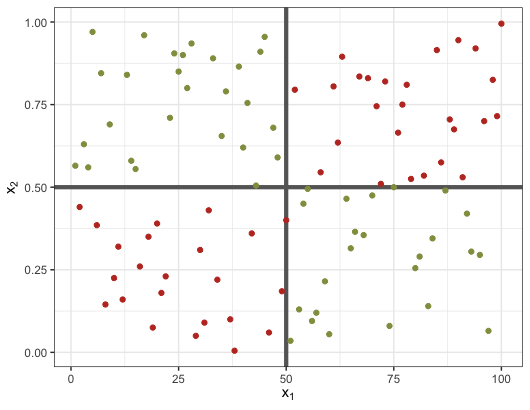}
\caption{Exemplification of the decision rules that would separate the two groups well.}
\label{fig:decision}
\end{figure}

There exist a few advantages of the use of trees. 
Non-linear decision regions between the covariables are
better captured by this algorithm than with usual 
regression models. Consider Figure \ref{fig:tree} 
as an example: a linear model would not be able
to reproduce the relationship between the two 
variables very well. As we see in Figure
\ref{fig:decision}, a tree model with the 
decision rules configured as the gray lines 
displayed, on the other hand,
would satisfactorily separate 
the observations in the two existing classes. 

The performance of the tree algorithm needs to be evaluated regarding its prediction power. A test set is generally 
used to produce the predictions for never seen data. 
The comparison is performed between what was observed 
for the test set and what the tree model predicts. 
The closer the predictions are to the vector of observations,
the better the algorithm is doing, and an accuracy measure is calculated as 

\begin{equation}
Acc = \frac{1}{n} \sum_{i = 1}^{n}
I(y_i \neq \hat y_i)
\end{equation}
where $I$ is the indicator for whether the model
prediction $\hat y_i$ is compatible with 
what was observed and $n$ is the sample size 
\cite{Hastie}.

Once we defined decision trees, we can extend the definition to the random forest algorithm. 
A random forest is an ensemble learning algorithm
introduced by Breiman \cite{Breiman2001}, 
that works by constructing a large number $B$ 
of trees and combining their results \cite{Hastie}. 
For each tree, a bootstrapped dataset 
\cite{Efron1979} and only a set of 
$m$ predictors, from the $p$ total, are considered. 
Typically, the variables are selected randomly for
each step of the algorithm and $m$ is chosen as 
approximately $\sqrt p$ \cite{Hastie}.   

Since the correlation between the trees can be an issue,
selecting just a few variables for each step is useful to
overcome this problem \cite{Bernard2010}.
If the trees are allowed to consider different variables at
each time, strong predictors will not dominate the algorithm,
as sometimes they will not be even available
for the split. Using this criterion, 
the results produced by the uncorrelated trees are more reliable \cite{Hastie}. 
The final prediction for the random forest 
classification algorithm 
is the \textit{majority vote} of the $B$ trees or, 
in other words, the classification proposed
by most of the trees is the one that will be
taken as the prediction of the final model.

\subsection{Genre Modeling}

The nine music genres present in our dataset 
were modeled with a random forest algorithm. Our target
variable is the music genres, while the predictors are 
the extracted and engineered features. In total, 
four models were fitted, in a nested fashion, structured
as

\begin{enumerate}
    \item First model: triads (6)
    \item Second model: triads (6) + tetrads (6)
    \item Third model: triads (6) + tetrads (6) + common transitions (3)
    \item Fourth model: triads (6) + tetrads (6) +
        common transitions (3) + miscellany (8).
\end{enumerate}

Our first model accounted for the first
six variables, related to the triad features of the chords. The following model was
added the tetrads variables and the third model
was built using also the common
transitions variables. Lastly, the fourth
model also considered the miscellany variables, 
adding up to 23 predictors. In this way, we could  
compare the predictions for all the different four models. 
The interest relied on assess the information brought 
by each thematic group of predictors. 

\section{RESULTS}

In this section, we present a summary
of the exploratory analysis along with
the results of the fitted models. The discussion
is shown in the following section.

\subsection{Exploratory Analysis}

\begin{figure}[t]
\centering
\includegraphics[width=0.75\columnwidth]{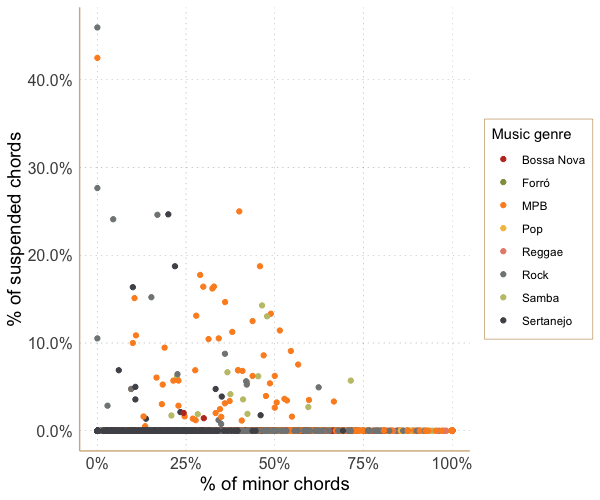}
\caption{Percentage of minor chords 
in the song \textit{versus} the percentage
of suspended chords, identifying the music
genre.}
\label{fig:sus}
\end{figure}

\begin{figure}[t]
\centering
\includegraphics[width=0.75\columnwidth]{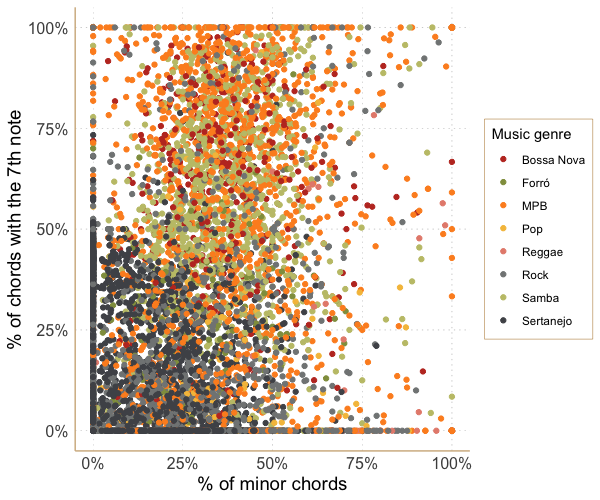}
\caption{Percentage of minor chords 
in the song \textit{versus} the percentage
of chords with the seventh note, 
identifying the music genre.}
\label{fig:7th}
\end{figure}

\begin{figure}[t]
\centering
\includegraphics[width=0.75\columnwidth]{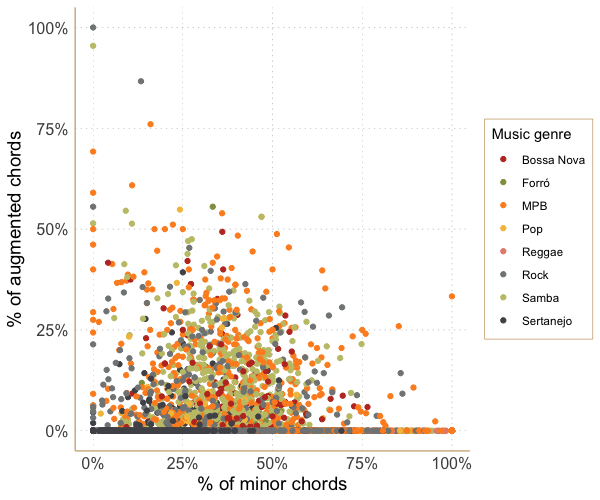}
\caption{Percentage of minor chords 
in the song \textit{versus} the percentage
of augmented chords, identifying the music
genre.}
\label{fig:aug}
\end{figure}

Non-linearity between predictor variables
is not an assumption of the random
forest model, but rather a problem that
the algorithm can overcome. Nevertheless,
Figures \ref{fig:sus}, \ref{fig:7th}
and \ref{fig:aug} illustrate the 
the relationship between some variables of
the first group. It is clear that
the relationships between the variables
follow a diversity of patterns between
the music genres. With the plots,
we can notice that the patterns 
would be poorly described if a linear
function was considered for this task,
as there is evidence of
non-linear relationships in our dataset.

\begin{figure}[h]
\centering
\includegraphics[width=0.8\columnwidth]{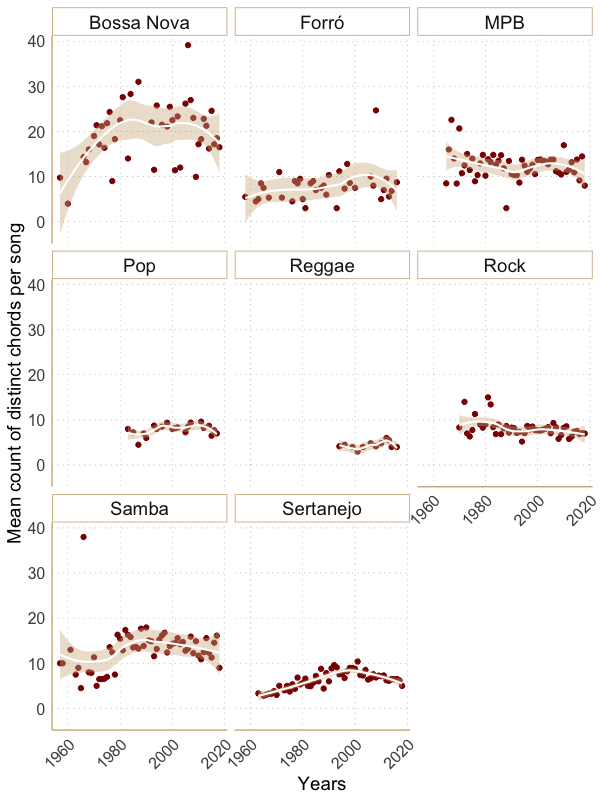}
\caption{Mean count of distinct chords in the songs for each year, by music genre. The white curve represents a local regression non-parametric smoother to highlight the trend of the observations.}
\label{fig:explor}
\end{figure}

In addition to that, it has become common sense to say 
that some music genres are harmonically simpler 
than others \cite{Almada2012}. We established as
a diversity measure the mean of the count of distinct 
chords, obtained per song for each genre. 
Figure \ref{fig:explor} shows how this mean behaves and
has changed over the years for the genres (1957 to 2018). Clearly, 
genres known as traditionally Brazilian, 
like Samba, MPB, and Bossa Nova, have higher values 
of the mean count of distinct chords. These genres 
also have a higher variation over time, 
which is in agreement with popular knowledge
\cite{Caldas2010}. Bossa Nova is the most
unstable one, having different values along the years
with an increasing trend, followed by Samba, that also
looks to have been going through similar changes. 
Sertanejo is the one that has the most evident
decrease, having a peak around 2000 and decay since that. The remaining genres are more harmonically uniform
and stable through time. We also observe
that data is not available for all of the years and genres, once some genres just appeared in Brazil in 
recent years \cite{Caldas2010}. 

\subsection{Random Forests Models}

The random forest model was chosen as the 
classification algorithm for the music genres. All 
of the final variables were used in the modeling. 
Our choice for this algorithm is justified by three main points:

\begin{itemize}
\item it easily allows the obtention of an 
importance measure of the predictors, 
\item there is no need to normalize or transform the variables, once the trees are invariant to the predictor's scale,
\item accommodation of non-linear relationships between the predictors and the response,
which are overcome by the model. 
\end{itemize}

The full dataset was randomly partitioned in train \thinspace (70\%) and test
\thinspace (30\%) set, with balancing by music
genre. The train set was used for the training of the model, while the test set serves for the calculation
of  performance measures in data not seen by the 
trained model. Table \ref{tab:descrip} presents 
the structure of the two datasets and 
the measures of goodness of fit of the four models
are in Table \ref{tab:qual}

\begin{table}[ht] \centering 
  \caption{Amount of songs for each genre, 
  in each partition, and representation (\%) of the genre in the full dataset.} 
  \label{} 
  \small
  \vspace{3mm}
\begin{tabular}{lllll} 
\\[-1.8ex]\hline 
\hline \\[-1.8ex] 
Genre & \multicolumn{1}{l}{Train} & \multicolumn{1}{l}{Test} &
\multicolumn{1}{l}{Representativity}  \\ 
\hline \\[-1.8ex] 
  Bossa Nova  & 305 (68\%)    &  133 (32\%)   & 438  (5.3\%)\\
  Forró       &  115 (73\%)   &  48  (27\%)   & 163  (2\%) \\
  MPB         & 1196 (67.8\%) & 476  (32.2\%) & 1679 (20.3\%)\\
  Pop         &  104 (66.4\%) &   39 (33.6\%) & 143  (1.7\%)\\
  Reggae      &  46 (68.1\%)  &   24 (31.9\%) & 70   (0.8\%)\\
  Rock        & 1127 (69.8\%) & 552  (30.2\%) & 1679 (20.4\%)\\
  Samba       & 877 (70.8\%)  & 378  (29.2\%) & 1255 (15.1\%)\\
  Sertanejo   & 1992 (68.2\%) & 849  (31.8\%) & 2841 (34.4\%)\\

\hline \hline  \\[-1.8ex] 
\end{tabular} 
\label{tab:descrip}
\end{table} 

\begin{table}[ht]
%\amll
\caption{Goodness of fit
for the four models: overall accuracy with lower and upper bounds and Kappa statistic with the respective p-value.}
\centering
\vspace{3mm}
\begin{tabular}{llllll}
\\[-1.8ex]\hline 
\hline \\[-1.8ex] 
 
Model & Accuracy & L.B. & U.B. & Kappa & P-Value \\ 
  \hline
  Model 1    & 0.53 & 0.51 & 0.55 & 0.37 & $<$ 0.0001 \\
  Model 2    & 0.57 & 0.54 & 0.59 & 0.42 & $<$ 0.0001 \\
  Model 3    & 0.59 & 0.56 & 0.60 & 0.44 & $<$ 0.0001 \\
  Model 4    & 0.62 & 0.60 & 0.64 & 0.49 & $<$ 0.0001 \\
\hline \hline  \\[-1.8ex] 
\end{tabular}
\label{tab:qual}

\end{table}

The \textit{Kappa} statistic shown in Table
\ref{tab:qual} is a metric of 
comparison between the observed accuracy
and the expected accuracy. Also known as
\textit{Non Information Rate}, or \textit{N.I.R},
the expected accuracy is the proportion of the 
most recurrent genre in the dataset, that in 
this case is 34\%.  The statistic is 
used to decide whether the classification
proposed by the model is more accurate 
than saying that all observations belong
to the most recurrent genre of the dataset
\cite{Cohen1960}. The formula
for the Kappa statistic is given 
as 

\begin{equation}
kappa = \frac{p_{0} - p_{e} }{1 - p_{e}},
\label{eq:kappa}
\end{equation}

where $p_{0}$ is the observed accuracy
and $p_{e}$ is the expected accuracy, and 
the \textit{Kappa} statistics follows an 
approximate Normal distribution asymptotically
\cite{Cohen1960}, which makes it easily 
testable. In the last column of Table \ref{tab:qual},
the p-values indicate whether 
the accuracy of the models is significantly
higher than the \textit{N.I.R}. We observe
that, for all different models, there are
evidence of their accuracy being 
significantly higher than using a naive
classification in the most common genre, 
or the non-information rate of the data.

From Table \ref{tab:qual}, we can also conclude
that the addition of new predictors 
sets progressively increased the accuracy
of the models. We need to remember that, in 
random forests models, the accuracy does not
behave as the $R^{2}$, used in the
evaluation of linear models \cite{rsquared}. The
$R^{2}$ invariably increases if we add more
variables to the model, once this measure
it's not penalized by the total number of 
parameters (in the case of a parametric model).
The random forest accuracy, on the other hand, 
can be negatively affected by the inclusion of
non-informative features that introduce noise, 
making the distinction between the classes a 
harder task \cite{Hastie}. 
Because of this property, it is safe to say that 
the variables added to the model are in fact making
the prediction capacity higher. 

The increase in the general accuracy is seemingly
uniform: to each new set of variables added, the 
increase is in about 3\%. The fourth model, will all 
the features, has a prediction capacity of around 62\%,
almost twice as the \textit{N.I.R.}. We conclude 
that genres of the songs present in our dataset can be
well predicted by looking at their harmonic features
as we extracted here. 

\begin{table}[ht]
\caption{Confusion matrix for the model
with only the  first set of variables.}
\small
\vspace{3mm}
 \setlength\tabcolsep{1.5pt}
\centering
\begin{tabular}{rrrrrrrrr}
 & B. Nova & Forró & MPB & Pop & Reggae & Rock & Samba & Sert. \\ 
  \hline
B. Nova   & \textbf{0.14} & 0.00 & 0.33 & 0.00 & 0.00 & 0.05 & 0.33 & 0.15 \\ 
  Forró      & 0.00 & \textbf{0.00} & 0.10 & 0.00 & 0.00 & 0.15 & 0.12 & 0.62 \\ 
  MPB        & 0.03 & 0.00 & \textbf{0.41} & 0.00 & 0.00 & 0.14 & 0.23 & 0.20 \\ 
  Pop        & 0.00 & 0.00 & 0.15 & \textbf{0.00} & 0.00 & 0.26 & 0.23 & 0.36 \\ 
  Reggae     & 0.00 & 0.00 & 0.25 & 0.00 & \textbf{0.00} & 0.50 & 0.04 & 0.21 \\ 
  Rock       & 0.01 & 0.00 & 0.11 & 0.00 & 0.00 & \textbf{0.34} & 0.07 & 0.47 \\ 
  Samba      & 0.02 & 0.00 & 0.26 & 0.00 & 0.00 & 0.05 & \textbf{0.57} & 0.11 \\ 
  Sert.  & 0.00 & 0.00 & 0.02 & 0.00 & 0.00 & 0.12 & 0.02 & \textbf{0.84} \\ 
   \hline
\end{tabular}
\label{tab:conf_1}
\end{table}

\begin{table}[ht]
\caption{Confusion matrix for the second model
with the first and second sets of variables.}
\small
\vspace{3mm}
 \setlength\tabcolsep{1.5pt}
\centering
\begin{tabular}{rrrrrrrrr}
 & B. Nova & Forró & MPB & Pop & Reggae & Rock & Samba & Sert. \\ 
  \hline
B. Nova  & \textbf{0.29} & 0.00 & 0.35 & 0.00 & 0.00 & 0.05 & 0.19 & 0.14 \\ 
  Forró     & 0.00 & \textbf{0.00} & 0.10 & 0.00 & 0.00 & 0.15 & 0.12 & 0.62 \\ 
  MPB       & 0.03 & 0.00 & \textbf{0.49} & 0.00 & 0.00 & 0.13 & 0.17 & 0.18 \\ 
  Pop       & 0.00 & 0.00 & 0.15 & \textbf{0.00} & 0.00 & 0.31 & 0.18 & 0.36 \\ 
  Reggae    & 0.00 & 0.00 & 0.17 & 0.00 & \textbf{0.00} & 0.50 & 0.12 & 0.21 \\ 
  Rock      & 0.00 & 0.00 & 0.13 & 0.00 & 0.00 & \textbf{0.36} & 0.06 & 0.44 \\ 
  Samba     & 0.02 & 0.00 & 0.20 & 0.00 & 0.00 & 0.04 & \textbf{0.63} & 0.10 \\ 
  Sert. & 0.00 & 0.00 & 0.02 & 0.00 & 0.00 & 0.12 & 0.02 & \textbf{0.84} \\
     \hline
\end{tabular}
\label{tab:conf_2}
\end{table}

\begin{table}[ht]
\caption{Confusion matrix for third model with the first, second and third sets of variables.}
\small
\vspace{3mm}
 \setlength\tabcolsep{1.5pt}
\centering
\begin{tabular}{rrrrrrrrr}
 & B. Nova & Forró & MPB & Pop & Reggae & Rock & Samba & Sert. \\ 
  \hline
B. Nova  & \textbf{0.29} & 0.00 & 0.35 & 0.00 & 0.00 & 0.05 & 0.17 & 0.13 \\ 
  Forró     & 0.00 & \textbf{0.00} & 0.06 & 0.00 & 0.00 & 0.21 & 0.08 & 0.65 \\ 
  MPB       & 0.03 & 0.00 & \textbf{0.55} & 0.00 & 0.00 & 0.12 & 0.15 & 0.15 \\ 
  Pop       & 0.00 & 0.00 & 0.23 & \textbf{0.00} & 0.00 & 0.13 & 0.21 & 0.44 \\ 
  Reggae    & 0.00 & 0.00 & 0.38 & 0.00 & \textbf{0.04} & 0.46 & 0.04 & 0.08 \\ 
  Rock      & 0.00 & 0.00 & 0.14 & 0.00 & 0.00 & \textbf{0.35} & 0.06 & 0.45 \\ 
  Samba     & 0.02 & 0.00 & 0.21 & 0.00 & 0.00 & 0.03 & \textbf{0.66} & 0.08 \\ 
  Sert. & 0.00 & 0.00 & 0.02 & 0.00 & 0.00 & 0.09 & 0.02 & \textbf{0.86} \\ 
     \hline
\end{tabular}
\label{tab:conf_3}
\end{table}

\begin{table}[ht]
\caption{Confusion matrix for the fourth model with all the considered variables.}
\small
\vspace{3mm}
 \setlength\tabcolsep{1.5pt}
\centering
\begin{tabular}{rrrrrrrrr}
 & B. Nova & Forró & MPB & Pop & Reggae & Rock & Samba & Sert. \\ 
  \hline
B. Nova  & \textbf{0.28} & 0.00 & 0.40 & 0.00 & 0.00 & 0.05 & 0.16 & 0.12 \\ 
  Forró     & 0.00 & \textbf{0.00} & 0.12 & 0.00 & 0.00 & 0.12 & 0.10 & 0.65 \\ 
  MPB       & 0.01 & 0.00 & \textbf{0.59} & 0.00 & 0.00 & 0.11 & 0.13 & 0.15 \\ 
  Pop       & 0.00 & 0.00 & 0.13 & \textbf{0.00} & 0.00 & 0.28 & 0.15 & 0.44 \\ 
  Reggae    & 0.00 & 0.00 & 0.25 & 0.00 & \textbf{0.08} & 0.46 & 0.08 & 0.12 \\ 
  Rock      & 0.00 & 0.00 & 0.16 & 0.00 & 0.00 & \textbf{0.43} & 0.05 & 0.35 \\ 
  Samba     & 0.01 & 0.00 & 0.20 & 0.00 & 0.00 & 0.03 & \textbf{0.66} & 0.10 \\ 
  Sert. & 0.00 & 0.00 & 0.02 & 0.00 & 0.00 & 0.07 & 0.02 & \textbf{0.89} \\

     \hline
\end{tabular}
\label{tab:conf_4}
\end{table}

We can also analyze the confusion matrices, 
that serve to show a detailed
report about how the classes of the factor variable
are being classified \cite{Hastie}. If the classes 
are being mistaken by a specific one, for example, this is 
the proper method to access this behavior. 
For each of the four models, we 
show the confusions matrices in Tables
\ref{tab:conf_1}, \ref{tab:conf_2}, \ref{tab:conf_3}  
and \ref{tab:conf_4}.

From Table \ref{tab:conf_1} to Table \ref{tab:conf_2},
there is a considerable increase on the correct 
classification rate, especially for Bossa Nova
(about 15\%), MPB (about 8\%) and Samba (about 6\%). 
This means that with the addition of the variables
about the tetrads, the most popular genres of Brazilian music are more easily identified by the model. In Table
\ref{tab:conf_3}, the increase occurs to MPB (about 6\%), 
Samba and Sertanejo (about 2\% for both), but it's more
intense for Reggae (4\%). On the previous models, this 
genre was being completely missed, but with the
information about the common chords transitions, at least 
some percentage of it is being distinguished, which 
is considered a significative information gain. Lastly,
on Table \ref{tab:conf_4}, the increase was specially high
for Rock (about 8\%), followed MPB and Reggae
(about 4\% for both) and Sertanejo (about 3\% of increase).
With that, we verify that the fourth set of variables, 
the miscellany that includes the popularity, year of the 
song and distances to the 'C' chord is notably relevant
in the classification of those genres, with emphasis 
on Rock.

\begin{figure}[h]
\centering
\includegraphics[width=0.9\columnwidth]{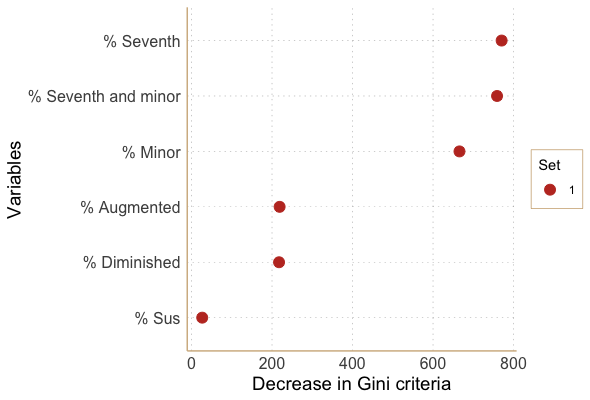}
\caption{Importance plot for the model
with only the first set of variables.}
\label{fig:imp0}
\end{figure}

\begin{figure}[h]
\centering
\includegraphics[width=0.9\columnwidth]{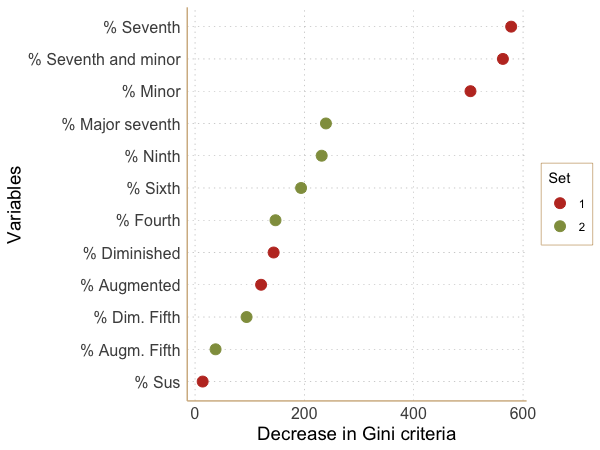}
\caption{Importance plot for the second model
with the first and second sets of variables.}
\label{fig:imp1}
\end{figure}

\begin{figure}[h]
\centering
\includegraphics[width=0.9\columnwidth]{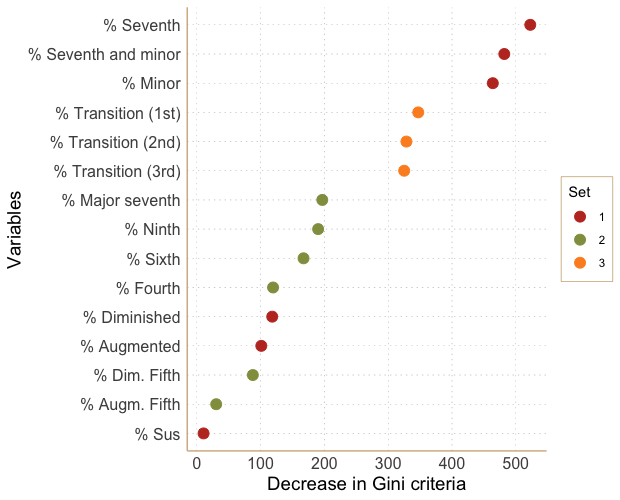}
\caption{Importance plot for third model with the first, second and third sets of variables.}
\label{fig:imp2}
\end{figure}

\begin{figure}[h]
\centering
\includegraphics[width=0.9\columnwidth]{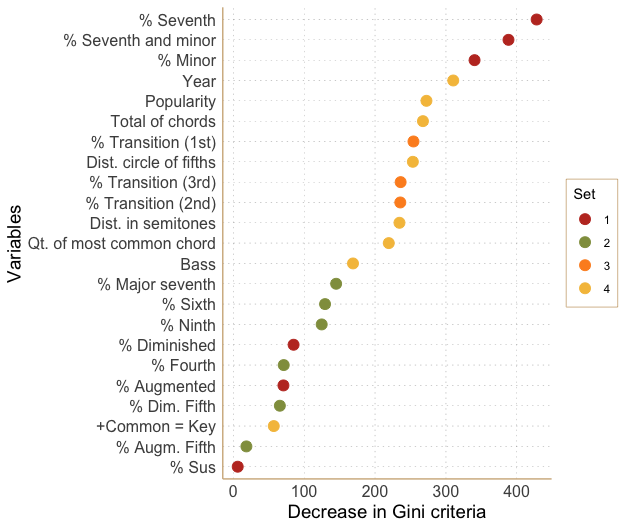}
\caption{Importance plot for the fourth model with all the considered variables.}
\label{fig:imp3}
\end{figure}

At last, we consider the specific importance of the 
features. An importance plot is used to show the measures
of how relevant to the model the features are,
by demonstrating the decrease in the Gini criteria
attributed to the splits in each variable \cite{Hastie}.
Therefore it is easy to perceive the prediction
strength of our variables, being that the strongest
ones will appear in the top of the plot. 
Given this and the plot being shown in 
Figure \ref{fig:imp0}, we notice that 
the first model has three main variables used in the correct
classification: the percentage of chords with the
seventh, the percentage of minor chords with the seventh. 
After the addition of the second set of variables, 
the same variables have remained on top (Figure \ref{fig:imp1}).
By that, we can infer that these three variables have higher discriminant power so far, unlike the second set of variables. 
The third set of variables, for which the addition
is reported on Figure \ref{fig:imp2},
have some strong variables, which are taking 
place of the old ones when added to the model.
This aspect shows that the features of the third
set carry a notorious amount of information about the 
response variables, our music genres. 
Finally, from the importance plot of the model 
with all of the variables, shown in Figure \ref{fig:imp3}, 
we conclude that the most relevant one out of the miscellany
set is the year of release of the song, followed by 
the popularity and the total number of chords in each song.
Those features took the place of the ones that were introduced
previously, which highlights how informative they are. 

In summary, the first set of variables is 
the most informative one. By using features
related to the triads and simple tetrads, the 
first features set makes it possible for us to separate 
the nine Brazilain genres, especially for the
traditional ones (Table \ref{tab:conf_1}). This means
that with the basic information about the songs we 
already obtained good results. The 
results improved mostly by the addition
of external variables, such as the year 
and popularity of each song, showing how 
the features from the streaming software 
Spotify carry relevant information
about the boundaries of the evaluated genres. 
Next, in the importance sequence, we have the transitions 
and distances variables, strengthening the idea of 
harmonic characteristics being important
to discriminate music genre and also clearing the 
path about what should be the further steps on the 
improvement of the analysis.

\section{CONCLUSIONS AND DISCUSSION}

With our results, we can conclude that is possible
to predict music genres of the Brazilian popular music
combining features extracted from their harmonic structures
and external variables.  The overall accuracy of the final 
model is 62\% with a confidence interval of [60\%, 64\%],
being that the better-classified genres are the Brazilian
Sertanejo, Samba, MPB (also known as Brazilian Popular Music) 
and Rock. The most discriminative variables for the model 
are the percentage of chords with the seventh note, 
percentage of minor chords with the seventh note, the percentage 
of  minor chords, the year of release of the songs, their
popularity, the behavior of the most common chord transitions,
the mean distances of the chords having the 'C' chord in
the circle of fifth as a reference and the total number of
chords in each song. On this group, prevail the 
features that can be extracted purely with the symbolic
part of the chords. The features obtained with the aid of
the Spotify API are also very influent, what makes sense 
once they came from a very influent music streaming software
and its data is reliable \cite{Schettino2017}. This 
accentuates how having access to data via the Spotify API is 
useful and pertinent to research in general. 

We emphasize that method for the obtention of 
the data is implemented in the \textit{chorrrds} 
package for R \cite{chorrrds}. If data for different artists
is required, the package would be the best the option
for that task. Moreover, every step of this analysis can be 
reproduced using the code available at: \\
\small\texttt{https://github.com/brunaw/genre\_classification }
It is very important for the authors that not only 
this work is entirely reproducible, but can also
be easily extended.

Next steps of this work include specially
the engineering of the new variables and 
applying different algorithms to model the data.
The modeling section can be improved in two
fundamental ways: changing the algorithm itself for
another suitable technique and improving the existent model. Options of possible new modeling techniques would be deep learning models \cite{Lecun2015} and
naive Bayes models \cite{Murphy2006}. 
References about similar approaches 
using deep learning can be found in \cite{cnnmusicgenre} and 
\cite{cnnmusicgenre2}. 

As for changing the existing model, 
alternatives are removing non-informative variables
and creating new ones. Using the importance criterion,
we have a very good direction about
which variables should be reconsidered to be kept
in the model. More sophisticated techniques such as
Principal Component Analysis \cite{Jolliffe2002}
could also play an important role in the task
of reducing the dimension of columns used in 
the model. Analogously, reducing the amount
of observations is considered a potential next step.
The idea here is to find the songs that carry most 
part of the variance for each genre and thereby 
can be useful to summarize them, allowing us to use less 
data to obtain similar results.  Regardless of that,
finding ways of engineering new variables is a key point 
of interest. As the main goal of this work is focused on 
showing how Brazilian music genres can be  predicted 
using interpretable engineered features, assessing more 
information about the songs via the creation of more features 
is essential.

\bibliographystyle{unsrt}  


\begin{thebibliography}{1}
\bibitem{Caldas2010}
Waldenyr Caldas.
\newblock {\em {Iniciação à Música Popular Brasileira}}, volume~1.
\newblock 2010.

\bibitem{Lena2008}
Jennifer~C. Lena and Richard~A. Peterson.
\newblock {Classification as culture: Types and trajectories of music genres}.
\newblock {\em American Sociological Review}, 2008.

\bibitem{Lee2004}
Jin~Ha Lee and J~Stephen Downie.
\newblock {Survey of music information needs, uses, and seeking behaviours:
  preliminary findings}, 2004.

\bibitem{Cheng2008}
{Automatic chord recognition for music classification and retrieval}.
\newblock In {\em 2008 IEEE International Conference on Multimedia and Expo,
  ICME 2008 - Proceedings}, 2008.

\bibitem{Correa2016}
D{\'{e}}bora~C. Corr{\^{e}}a and Francisco~Ap Rodrigues.
\newblock {A survey on symbolic data-based music genre classification}, 2016.

\bibitem{albin2003livro}
R.C. Albin.
\newblock {\em O livro de ouro da MPB: a hist{\'o}ria de nossa m{\'u}sica
  popular de sua origem at{\'e} hoje}.
\newblock Livros de ouro. Ediouro, 2003.

\bibitem{Burgoyne2015}
John~Ashley Burgoyne, Ichiro Fujinaga, and J.~{Stephen Downie}.
\newblock {Music Information Retrieval}.
\newblock In {\em A New Companion to Digital Humanities}. 2015.

\bibitem{Pampalk}
Elias Pampalk, Arthur Flexer, and Gerhard Widmer.
\newblock {Improvements of Audio-Based music similarity and genre
  classification}.
\newblock In {\em ISMIR}, 2005.

\bibitem{Tzanetakis}
George Tzanetakis and Perry Cook.
\newblock {Musical genre classification of audio signals}.
\newblock {\em IEEE Transactions on Speech and Audio Processing}, 2002.

\bibitem{Bahuleyan}
Hareesh Bahuleyan.
\newblock {Music Genre Classification using Machine Learning Techniques}.
\newblock 2018.

\bibitem{Scaringella}
Nicolas Scaringella, Giorgio Zoia, and Daniel Mlynek.
\newblock {Automatic genre classification of music content}.
\newblock {\em IEEE Signal Processing Magazine}, 2006.

\bibitem{Neuman2016}
Yair Neuman, Leonid Perlovsky, Yochai Cohen, and Danny Livshits.
\newblock {The personality of music genres}.
\newblock {\em Psychology of Music}, 2016.

\bibitem{Chuan2018}
Ching-Hua Chuan, Kat Agres, and Dorien Herremans.
\newblock {From Context to Concept: Exploring Semantic Relationships in Music
  with Word2Vec}.
\newblock nov 2018.

\bibitem{PerezSancho2010}
Carlos P{\'{e}}rez-Sancho, David Rizo, Jos{\'{e}}~M. Iesta, Pedro~J.Ponce {De
  Le{\'{o}}n}, Stefan Kersten, and Rafael Ramirez.
\newblock {Genre classification of music by tonal harmony}.
\newblock {\em Intelligent Data Analysis}, 2010.

\bibitem{Vianna1999}
{\em {The mystery of samba : popular music {\&} national identity in Brazil}}.
\newblock 1999.

\bibitem{Behague1980}
Gerard B{\'{e}}hague.
\newblock {Brazilian Musical Values of the 1960s and 1970s: Popular Urban Music
  from Bossa Nova to Tropicalia}.
\newblock {\em The Journal of Popular Culture}, 1980.

\bibitem{McCann2007}
Bryan McCann.
\newblock {Blues and Samba: Another Side of Bossa Nova History.}
\newblock {\em Luso-Brazilian Review}, 2007.

\bibitem{Muller2007}
Meinard M{\"{u}}ller.
\newblock {\em {Information retrieval for music and motion}}.
\newblock 2007.

\bibitem{Almada2012}
Carlos Almada.
\newblock {\em {Harmonia Funcional}}, volume~1.
\newblock 2012.

\bibitem{Rowe2005}
Robert Rowe.
\newblock {:The Cognition of Basic Musical Structures}.
\newblock {\em Music Perception}, 2005.

\bibitem{Iacus2015}
Stefano~M. Iacus.
\newblock {Automated Data Collection with R - A Practical Guide to Web Scraping
  and Text Mining}.
\newblock {\em Journal of Statistical Software}, 2015.

\bibitem{Schettino2017}
Vinicius J.~Schettino, Regina Braga, José Maria N.~David, and Marco Antônio
  P.~Araújo.
\newblock In {\em Proceedings of the 11th Brazilian Symposium on Software
  Components, Architectures, and Reuse - SBCARS '17}.

\bibitem{chorrrds}
Bruna Wundervald.
\newblock The chorrrds package for extraction of music chords data in r, 2018.

\bibitem{Rsoftware}
{R Core Team}.
\newblock {\em R: A Language and Environment for Statistical Computing}.
\newblock R Foundation for Statistical Computing, Vienna, Austria, 2018.

\bibitem{featureeng}
M.~Kuhn and K.~Johnson.
\newblock {\em Feature Engineering and Selection: A Practical Approach for
  Predictive Models}.
\newblock 2018.

\bibitem{nargesian2017learning}
Fatemeh Nargesian, Horst Samulowitz, Udayan Khurana, Elias~B Khalil, and Deepak
  Turaga.
\newblock Learning feature engineering for classification.
\newblock In {\em Proceedings of the 26th International Joint Conference on
  Artificial Intelligence}, pages 2529--2535. AAAI Press, 2017.

\bibitem{Lewin1982}
D~Lewin.
\newblock {A formal theory of generalized tonal functions}.
\newblock {\em Journal of Music Theory}, 1982.

\bibitem{Lerdahl1977}
{Toward a Formal Theory of Tonal Music}.
\newblock {\em Journal of Music Theory}, 1977.

\bibitem{Absolu2010}
Brandt Absolu, Tao Li, and Mitsunori Ogihara.
\newblock {Analysis of chord progression data}.
\newblock {\em Studies in Computational Intelligence}, 2010.

\bibitem{Faraway2016}
J~J Faraway.
\newblock {\em {Extending the Linear Model with R: Generalized Linear, Mixed
  Effects and Nonparametric Regression Models}}.
\newblock Chapman {\&} Hall/CRC Texts in Statistical Science. CRC Press, 2016.

\bibitem{Kingsford2008}
Carl Kingsford and Steven~L Salzberg.
\newblock {What are decision trees?}
\newblock {\em Nature Biotechnology}, 2008.

\bibitem{Hastie}
Jerome {Hastie, Trevor, Tibshirani, Robert, Friedman}.
\newblock {\em {The Elements of Statistical Learning, Second Edition}}.
\newblock 2009.

\bibitem{Breiman2001}
Leo Breiman.
\newblock {Random forests}.
\newblock {\em Machine Learning}, 2001.

\bibitem{Efron1979}
B.~Efron.
\newblock {Bootstrap Methods: Another Look at the Jackknife}.
\newblock {\em The Annals of Statistics}, 1979.

\bibitem{Bernard2010}
Simon Bernard, Laurent Heutte, and S{\'{e}}bastien Adam.
\newblock {A study of strength and correlation in random forests}.
\newblock In {\em Communications in Computer and Information Science}, 2010.

\bibitem{Cohen1960}
Jacob Cohen.
\newblock {A Coefficient of Agreement for Nominal Scales}.
\newblock {\em Educational and Psychological Measurement}, 1960.

\bibitem{rsquared}
Robert G. D. (Robert George~Douglas) Steel and 1908 Torrie, James H.
  (James~Hiram).
\newblock {\em Principles and procedures of statistics : with special reference
  to the biological sciences}.
\newblock New York : McGraw-Hill, 1960.
\newblock Includes bibliographical references.

\bibitem{Lecun2015}
Yann Lecun, Yoshua Bengio, and Geoffrey Hinton.
\newblock {Deep learning}, 2015.

\bibitem{Murphy2006}
Kevin~P Murphy.
\newblock {Naive Bayes classifiers}.
\newblock {\em Bernoulli}, 2006.

\bibitem{cnnmusicgenre}
Hareesh Bahuleyan.
\newblock Music genre classification using machine learning techniques.
\newblock {\em CoRR}, abs/1804.01149, 2018.

\bibitem{cnnmusicgenre2}
Sergio Oramas, Oriol Nieto, Francesco Barbieri, and Xavier Serra.
\newblock Multi-label music genre classification from audio, text, and images
  using deep features.
\newblock {\em CoRR}, abs/1707.04916, 2017.

\bibitem{Jolliffe2002}
I~T Jolliffe.
\newblock {\em {Principal Component Analysis. Second Edition}}.
\newblock 2002.


\end{thebibliography}

\end{document}